\documentclass[twocolumn,trackchanges]{aastex631}

\usepackage[utf8]{inputenc}
\usepackage{xcolor}
\usepackage{hyperref}
\usepackage{amsmath,units}
\usepackage{xspace}
\usepackage{acronym}
\usepackage{url}
\usepackage{placeins}

\newcommand{\result}[1]{{#1}}
\newcommand{\chase}[1]{{#1}}

\newcommand{\Msun}{\ensuremath{M_{\odot}}}
\newcommand{\MbhRLO}{\ensuremath{M_{\mathrm{1,RLO2}}}}
\newcommand{\MdonRLO}{\ensuremath{M_{\mathrm{2,RLO2}}}}
\newcommand{\PRLO}{\ensuremath{P_{\mathrm{RLO2}}}}
\newcommand{\MbhZams}{\ensuremath{M_{\mathrm{1,ZAMS}}}}
\newcommand{\MdonZams}{\ensuremath{M_{\mathrm{2,ZAMS}}}}

\newcommand{\MdonPre}{\ensuremath{M_{\mathrm{2,preSN}}}}

\newcommand{\VpecBirth}{\ensuremath{V_{\mathrm{pec,birth}}}}
\newcommand{\VpecBirthObs}{\ensuremath{V_{\mathrm{pec,birth,obs}}}}
\newcommand{\qObs}{\ensuremath{q_{\mathrm{obs}}}}
\newcommand{\MbhObs}{\ensuremath{M_{\mathrm{1,obs}}}}
\newcommand{\MdonObs}{\ensuremath{M_{\mathrm{2,obs}}}}

\newcommand{\Vkick}{\ensuremath{V_{\mathrm{kick}}}}

\newcommand{\TeffObs}{\ensuremath{T_{\mathrm{eff,obs}}}}
\newcommand{\posydon}{\texttt{POSYDON}}
\newcommand{\sys}{J1305}
\newcommand{\system}{MAXI J1305-704} 
\newcommand{\CIERA}{Center for Interdisciplinary Exploration and Research in Astrophysics (CIERA), Department of Physics and Astronomy, Northwestern University, 1800 Sherman Avenue, Evanston, IL 60201, USA}
\newcommand{\BU}{Department of Astronomy, Boston University, 725 Commonwealth Avenue Boston, MA 02215, USA}
\def\MbhMed{\ensuremath{8.9}}

\def\MbhHigh{\ensuremath{12.1}}
\def\MbhFull{\ensuremath{\MbhMed_{-1.0}^{+1.6}}}
\def\MdonMed{\ensuremath{0.43}}

\def\MdonFull{\ensuremath{\MdonMed_{-0.16}^{+0.16}}}
\def\TeffMed{\ensuremath{4610}}

\def\TeffFull{\ensuremath{\TeffMed_{-160}^{+130}}}
\def\qMed{\ensuremath{0.05}}

\def\qFull{\ensuremath{\qMed_{-0.02}^{+0.02}}}
\def\PMed{\ensuremath{0.394}}

\def\PFull{\ensuremath{\PMed_{-0.004}^{+0.004}}}

\def\pmra{\ensuremath{-7.89}}
\def\sigpmra{\ensuremath{0.62}}
\def\pmdec{\ensuremath{-0.16}}
\def\sigpmdec{\ensuremath{0.72}}

\def\sysX{\ensuremath{-3.87}}
\def\sysY{\ensuremath{-6.15}}
\def\sysZ{\ensuremath{-0.98}}
\def\sysVx{\ensuremath{-227}}
\def\sysVz{\ensuremath{20}}
\def\sysVy{\ensuremath{-140}}

\def\VkickLowwAll{\ensuremath{70}}
\def\VkickMedwAll{\ensuremath{91}}

\def\VkickPluswAll{\ensuremath{23}}
\def\VkickMinuswAll{\ensuremath{12}}

\def\VblaauwMedwAll{\ensuremath{0.7}}

\def\VblaauwPluswAll{\ensuremath{1.0}}
\def\VblaauwMinuswAll{\ensuremath{0.5}}

\def\dMHighwAll{\ensuremath{0.9}}

\begin{document}

\title{A Black Hole Kicked At Birth: \system{}}

\author[0000-0001-9879-6884]{Chase~Kimball}
\email{CharlesKimball2022@u.northwestern.edu}
\affiliation{\CIERA}

\author[0000-0002-6269-0970]{Sam~Imperato}
\affiliation{\CIERA}
\affiliation{\BU}

\author[0000-0001-9236-5469]{Vicky~Kalogera}
\affiliation{\CIERA}

\author[0000-0003-4474-6528]{Kyle\,A.\,Rocha}
\affiliation{Center for Interdisciplinary Exploration and Research in Astrophysics (CIERA), Northwestern University, 2145 Sheridan Road,
Evanston, IL 60208, USA}
\affiliation{Department of Physics and Astronomy, Northwestern University, 2145 Sheridan Road, Evanston, IL 60208, USA}

\author[0000-0002-2077-4914]{Zoheyr Doctor}
\affiliation{ \CIERA{} }

\author[0000-0001-5261-3923]{Jeff~J.~Andrews}
\affiliation{ \CIERA{} }
\affiliation{Department of Physics, University of Florida, 2001 Museum Rd, Gainesville, FL 32611, USA}

\author[0000-0002-4442-5700]{Aaron\,Dotter}
\affiliation{ \CIERA{} }

\author[0000-0002-7464-498X]{Emmanouil\,Zapartas}
\affiliation{Département d’Astronomie, Université de Genève, Chemin Pegasi 51, CH-1290 Versoix, Switzerland}
\affiliation{IAASARS, National Observatory of Athens, Vas. Pavlou and I. Metaxa, Penteli, 15236, Greece}

\author[0000-0002-3439-0321]{Simone\,S.\,Bavera}
\affiliation{Département d’Astronomie, Université de Genève, Chemin Pegasi 51, CH-1290 Versoix, Switzerland}

\author[0000-0003-3684-964X]{Konstantinos\,Kovlakas}
\affiliation{Département d’Astronomie, Université de Genève, Chemin Pegasi 51, CH-1290 Versoix, Switzerland}

\author[0000-0003-1474-1523]{Tassos\,Fragos}
\affiliation{Département d’Astronomie, Université de Genève, Chemin Pegasi 51, CH-1290 Versoix, Switzerland}

\author{Philipp\,M.\,Srivastava}
\affiliation{Center for Interdisciplinary Exploration and Research in Astrophysics (CIERA), Northwestern University, 2145 Sheridan Road,
Evanston, IL 60208, USA}
\affiliation{Electrical and Computer Engineering, Northwestern University, 2145 Sheridan Road, Evanston, IL 60208, USA}

\author[0000-0003-4260-960X]{Devina\,Misra}
\affiliation{Département d’Astronomie, Université de Genève, Chemin Pegasi 51, CH-1290 Versoix, Switzerland}

\author[0000-0001-9037-6180]{Meng\,Sun}
\affiliation{Center for Interdisciplinary Exploration and Research in Astrophysics (CIERA), Northwestern University, 2145 Sheridan Road,
Evanston, IL 60208, USA}

\author[0000-0002-0031-3029]{Zepei\,Xing}
\affiliation{Département d’Astronomie, Université de Genève, Chemin Pegasi 51, CH-1290 Versoix, Switzerland}

\begin{abstract}
When a compact object is formed in a binary, any mass lost during core collapse will impart a kick on the binary's center of mass. Asymmetries in this mass loss or neutrino emmission would impart an additional natal kick on the remnant black hole or neutron star, whether it was formed in a binary or in isolation. While it is well established that neutron stars receive natal kicks upon formation, it is unclear whether black holes do as well. Here, we consider the low-mass X-ray binary \system{}, which has been reported to have a space velocity $\gtrsim$ 200 km/s. In addition to integrating its trajectory to infer its velocity upon formation of its black hole, \chase{we account for recent estimates of its period, black hole mass, mass ratio, and donor effective temperature from photometric and spectroscopic observations. We find that if \system{} formed via isolated binary evolution in the thick Galactic disk, \result{then the supernova that formed its black hole imparted a natal kick of at least 70 km/s while ejecting less than $\simeq 1$ M$_\odot$ with 95\% confidence assuming uninformative priors on mass loss and natal kick velocity.}}
\end{abstract}
\keywords{
Astrophysical black holes, Binary evolution, Supernovae, Natal kicks
}

\section{Introduction}
\label{sec:intro}

The physics underlying supernovae (SNe) and explosive mass loss may be imprinted on the velocities of  the compact objects they leave behind. While any sudden mass lost from a component of a binary system will impart a velocity on the center of mass proportional to the amount of mass lost \citep{Blaauw}, the compact object remnants themselves may also receive an additional ``natal" kick on the order of hundreds of kilometers per second due to asymmetries in the SN mechanism \citep{Shklovskii,Janka2012,Wongwathanarat}. Surveys of the proper motions and heights above the Galactic disk \citep{Lyne1994,Hobbs2005,Kapil2022} have strongly suggested that neutron star remnants must experience these kicks, as have binary evolutionary analyses (e.g. \citet{Wong2010,Tauris2017} and references therein). However, the evidence that black holes (BHs) receive natal kicks is less clear.

In recent years, aided by the growing wealth of BH observations, studies have worked toward constraining BH natal kicks using a variety of methods and data sets. A number of studies have focused on the population of massive runaway stars with large space velocities, considering the possibility that they may have originated in binaries that were disrupted via the core collapse of their companions, possibly with the assistance of natal kicks \citep{vanRensbergen1996, deDonder1997,Dray2005, Eldridge2011,Boubert2018,Renzo2019,Aghakhanloo2022}. 

Other studies have turned to X-ray binaries (XRBs) with BH accretors --  aided by the growing subset of systems with known proper motions and radial velocities --- to constrain natal BH kicks. Some used just these kinematic constraints  to estimate birth velocities of known XRBs, suggesting that they may have received kicks in excess of what they could have received due to symmetric mass loss alone \citep{Mirabel2001,Repetto2012,Atri,Sanchez2021}. Others combined the kinematic constraints with observational constraints on the orbital, BH, and donor properties, modeling the binary evolution and core collapse of their XRB progenitors to individually constrain mass loss and natal kicks \citep{Willems2005,Fragos2009, Wong2012, Wong2014}. Of these studies that considered both kinematic and binary/stellar observational constraints, most found only upper limits on the BH natal kicks with one exception: XTE J1118+40, which \citet{Fragos2009} found must have received a natal kick in excess of $\simeq$80 km/s. Here, we focus on the case of the low-mass X-ray binary (LMXB) \system{} because of its high peculiar velocity ($V_{\mathrm{pec}} \simeq 80_{-30}^{+30}$ km/s), distance above the Galactic plane ($|Z|\simeq$ 1 kpc), and short orbital period (P$\simeq 0.4$ days), reported in  \citet{Sanchez2021}. We \result{constrain} both its evolutionary and kinematic history to find a lower limit on its BH natal kick, making it only the second fully-modeled BH system that must have received a natal kick at birth.

While BH LMXBs may form dynamically via binary exchanges or single-single captures \citep{Clark1975,Hills1976}, here we work under the assumption that \system{} (hereafter \sys{}) formed in the Galactic field and evolved as an isolated binary. The standard isolated formation channel for BH LMXBs usually starts with an unequal-mass zero-age main sequence binary with a primary $\gtrsim 20$ \Msun that fills its Roche lobe (RL), undergoing a period of stable or unstable mass transfer (MT) -- the latter case leading to a common envelope (CE) phase. If the binary survives without merging first, the primary is left as a massive stripped naked helium star in a tight, circularized, orbit with a relatively unevolved companion. Upon core collapse, the orbit may be altered or completely disrupted due to mass ejection and a possible natal kick. If it survives, the resulting detached binary will evolve until the secondary fills its Roche lobe and begins transferring mass onto the primary BH, forming a BH LMXB. In the case of \sys{}, with a sub-solar donor mass with a likely convective envelope, orbiting its BH with a sub-day period, the onset of RLO may have been aided by magnetic braking, which would have counteracted the tendency of the orbit to expand via mass transfer from the lighter secondary onto the heavier BH \citep{Konstantin2014}.

In this study, we combine forward-modeling of this evolutionary channel with backward modeling of the kinematic history of \sys{} to constrain the range of natal kicks its BH may have experienced -- following a method developed by \citet{Willems2005}, \citet{Fragos2009}, and \citet{Wong2012,Wong2014}. \result{While the kinematic history gives us constraints on the systemic velocity of the system at the birth of the BH, the forward modeling allows us to extract any asymmetric natal kick -- in addition to the Blaauw kick from mass-loss alone -- needed to attain that velocity.} \result{We start by evolving \sys{}'s trajectory backwards through the Galactic potential to construct a distribution of potential peculiar velocities of the system at the birth of its BH. We then run a series of detailed mass transfer sequences using \texttt{MESA}\footnote{Revision  11701 and the 20190503  version  of  the MESA software  development  kit } \citep{Paxton2010,Paxton2013,Paxton2015,Paxton2019} from the RLO of the secondary (RLO2) forwards, identifying potential progenitor binaries that simultaneously satisfy the observations of \sys{}'s BH mass, mass ratio, donor effective temperature, and orbital period. We then use \posydon{} \citep{Posydon} to constrain the parameter space of the possible progenitors that could produce these RLO2 binaries.}

\result{We note that when modeling core collapse, we choose not to adopt a SN prescription and instead sample mass-loss and  natal kick velocity from uniform, uninformative priors. Modern SN prescriptions typically scale BH kicks according to ejecta mass which is calculated via fitting formulae (see, e.g. \citet{Bray2018, Giacobbo2020} and references therein). However, BH kicks and the mechanisms that produce them are poorly understood, and here we aim to \textit{infer} magnitude of SN mass loss and BH kicks under uninformative priors, potentially informing SN prescriptions.}

\chase{Using our kinematic constraint on the total birth velocity of \sys{}, we then extract the range of required natal BH kicks from our forward modeling, finding that \sys{}'s BH likely received a natal kick greater than 70 km/s at 95\% confidence. In Section \ref{sec:observations}, we discuss the observational constraints on the stellar, binary, and kinematic properties of \sys{}. In Section \ref{sec:kinematics} and \ref{sec:evolution}, we describe how we reconstruct its kinematic history and constrain its possible pre- and post-SN properties, combining the results in \ref{sec:results}. We discuss our conclusions in Section 
\ref{sec:conclusions}}.

\section{\system{}}
\label{sec:observations}
While there are hundreds of known XRB sources, only about two dozen have confirmed BH accretors. Of these, 21 are classified as LMXBs \citep{Jonker2021}, having companions less than a few solar masses. There are proper motion measurements in the literature for 16 of these, and only 12 are complete with radial velocity measurements \citep{Atri}, crucial to estimating peculiar velocities and inferring BH kicks. Of these systems, \sys{} is particularly interesting because of its high distance above the Galactic plane, with $|Z|\approx$ 1 kpc,  and low orbital period ($\approx 0.4$ days), which is particularly helpful in constraining its progenitor properties (see Section \ref{sec:evolution} for details).
\sys{} was first discovered by the International Space Station's Monitor of All-sky X-ray Image instrument \citep{MAXI, Mihara2011}, and was first identified as an X-ray transient in \citet{Sato2012}. Follow-up observations classified \sys{} as an LMXB, potentially accreting onto a stellar-mass BH \citep{Greiner2012,Kennea2012,Suwa2012,Morihana2013}. The nature of its accretor was confirmed by \citet{Sanchez2021}, after photometric and spectroscopic observations of \sys{} in quiescence. They found that \sys{} has an orbital period of \PFull days, consisting of a black hole of mass \MbhObs$ = \MbhFull$ \Msun{}. With an observed mass ratio \qObs$=\qFull$ and donor effective temperature \TeffObs$=\TeffFull$ K, they find that the companion is an evolved dwarf star with \MdonObs$ = \MdonFull$ \Msun{} and an effective temperature of $T_{\mathrm{eff}}=\TeffFull$ K. They estimated that \sys{} is at a distance of $d=7.5_{-1.4}^{+2.8}$ kpc, with a radial velocity of $\gamma=9_{-5}^{+5}$ km/s. Combining this together with proper motion measurements in the direction of right ascension and declination ($\alpha$ and $\delta$, respectively) from GAIA \citep{GAIADR3} of $\Delta \alpha \cos{\delta} = \pmra_{-\sigpmra}^{+\sigpmra}$ mas yr$^{-1}$ and $\Delta \delta = \pmdec_{-\sigpmdec}^{+\sigpmdec}$, they calculated a peculiar velocity with respect to the local Galactic rotation of $V_{\mathrm{pec}} = 80_{-30}^{+30}$ km/s.  While we can not exclude that \sys{} originated in a globular cluster, this suggests that if it indeed originated in the thick disk, then \sys{} may have received a natal kick at birth.

\section{Constraining the progenitor of \sys{}}
\label{sec:progenitor}
\chase{We mostly follow the methodology described in \citet{Wong2014}, using kinematic modeling to infer the potential peculiar velocities of \sys{} at the birth of its black hole, and then constrain the possible pre- and post-SN binary properties using forward modeling to extract the asymmetric natal kick on the black hole from the total imparted peculiar velocity inferred kinematically. However, rather than use a rapid population synthesis code based on fits to single-star models when evolving potential progenitors, we compute detailed binary stellar evolution sequences using \texttt{MESA}, and generate populations with \posydon{} \citep{Posydon}, which is trained on \texttt{MESA} binary evolution grids and fully consistent with its treatment of binary stellar evolution and mass transfer. Except for where otherwise noted in relation to our treatment of core collapse, all of our parameter choices and prescriptions are as described in \posydon{} v1.}

\subsection{Kinematics}
\label{sec:kinematics}

\begin{figure*}
\includegraphics[width=\textwidth]{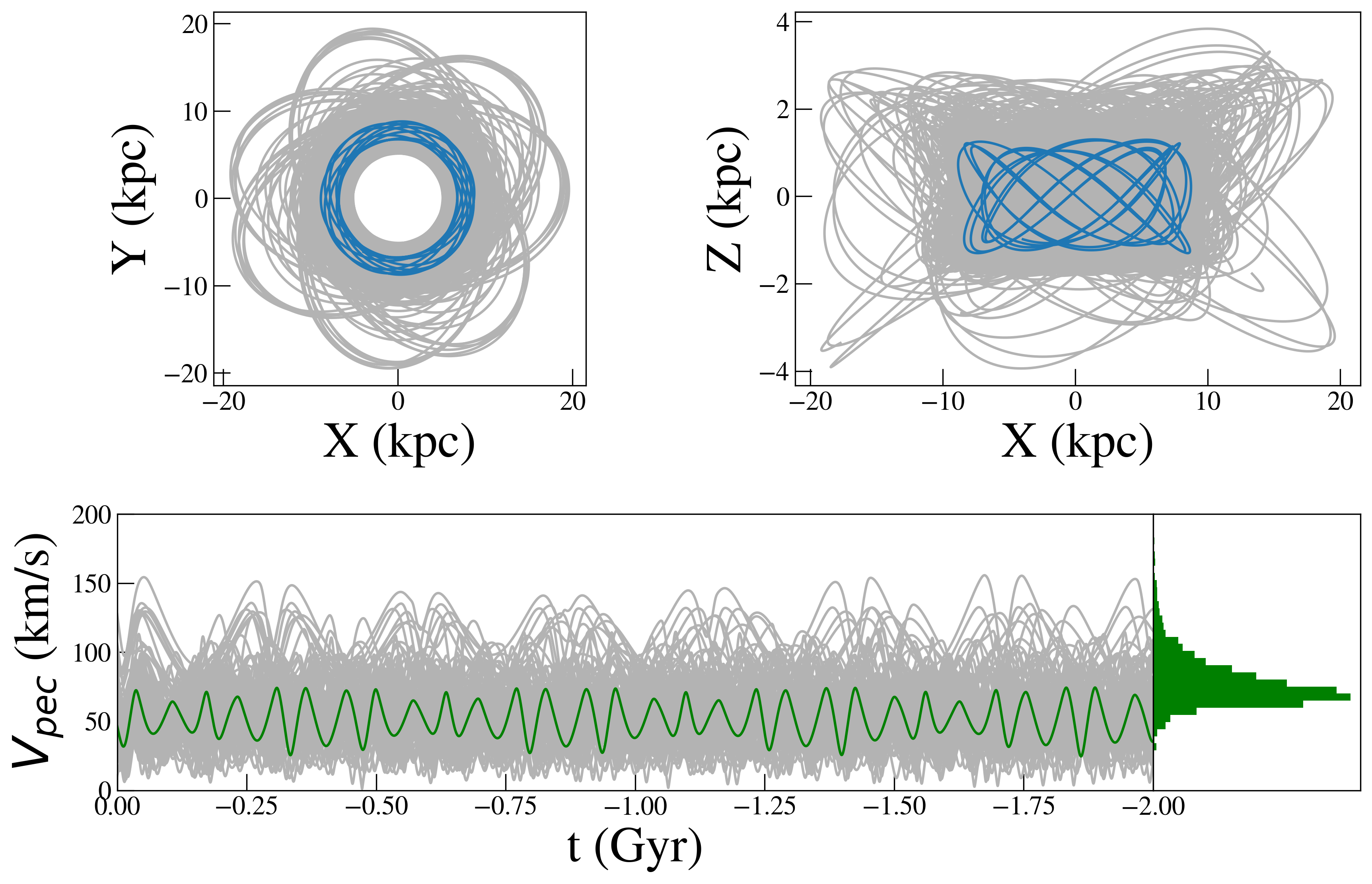}
\caption{Here we plot the possible kinematic histories of \sys{}. In blue we plot the trajectory corresponding to taking the median observed position and proper motion as initial conditions. In grey we draw random initial conditions from the corresponding posteriors. In the top two panels, we plot its trajectory in the X-Y and X-Z planes in Galactocentric coordinates. In the bottom panel, we plot a segment of the peculiar velocity as a function of time before the present. The rotated histogram on the right is the distribution of potential peculiar velocities at birth, inferred by sampling the trajectory at crossings of the Galactic plane that coincide (to within 10 Myrs) with donor ages from our successful RLO2 binaries.}
\label{fig:kinematics}
\end{figure*}
\chase{We begin by using the observed proper motion of \sys{} to constrain the total velocity imparted by the SN upon the birth of its BH.} Following the methodology of \citet{Wong2014}, we assume that \sys{} formed in the Galactic disk and integrate its trajectory backwards to infer the systemic velocity of \sys{} upon the birth of its black hole (\VpecBirthObs). \chase{In Section \ref{sec:evolution}, we infer the possible range of pre- and post-SN properties of the binary to extract the range of natal kicks required to create this birth velocity.}

We model the Milky Way with the static potential of \cite{Carlberg1987} and updated parameters from \cite{Kuijken1989}. We start with a coordinate system with axes XYZ with origin at the Galactic center, where Z=0 coincides with the Galactic mid-plane, the positive Y axis is in the direction of Galactic rotation at the location of the Sun, and the negative X-axis connects the projection of the Sun onto Galactic midplane to the Galactic center. Assuming a distance from the Sun to the Galactic center of $R_0 = $ 8.05 kpc and local rotational velocity of $\Omega_0=$ 238 km/s \cite{Honma2012}, this places the Sun at (X$_\odot$,Y$_\odot$,Z$_\odot$) = (-8.05, 0, .03) kpc with a velocity of (U$_\odot$,V$_\odot$,W$_\odot$) = (11.1, 12.24, 7.25) km/s with respect to the local standard of rest. Meanwhile, adopting the distance to \sys{} of $d=7.5_{-1.4}^{+2.8}$ kpc from the analysis in \citet{Sanchez2021}, this puts \sys{} at (X,Y,Z) = (\sysX, \sysY, \sysZ) kpc. Combining this with the radial velocity $\gamma=9_{-5}^{+5}$ km/s from that analysis with the angular velocity measurements ($\Delta \alpha$, $\Delta \delta$) = ($\pmra_{-\sigpmra}^{+\sigpmra}$,$\pmdec_{-\sigpmdec}^{+\sigpmdec}$ mas yr$^{-1}$ from the GAIA Early Data Release 3 \citep{GAIADR3}, we compute the proper motion of \sys{} with respect to our local standard of rest, finding (U,V,W) = (\sysVx, \sysVy, \sysVz) km/s.

In order to estimate \sys's peculiar velocity upon the birth of its black hole, we integrate its trajectory backwards in time, and sample potential peculiar velocities at moments that coincide with crossings of the Galactic mid-plane within 10 Myr of the age of a donor star from the winning RLO2 sequences. We take the donor age as a proxy for black hole age under the assumption that the lifetime of the black hole progenitor is negligible with respect to the age of our winning donors. We plot the result in Figure \ref{fig:kinematics}, finding that \sys{} likely had a peculiar velocity of \VpecBirthObs $=74_{-11}^{+19}$ km/s just after the birth of its black hole, and treat this as an observable. Assuming that prior to  the progenitor to \sys{} was moving in the Galactic mid-plane with the local Galactic rotation, this gives a measure of the total velocity imparted on \sys's center of mass upon core-collapse. This assumption that \sys{} was born directly in the disk at Z = 0 kpc with exactly the circular velocity at that location equates to assuming a ``kinematically cold" disk. To check that our results are not affected by this assumption, we compared \sys{}'s current peculiar velocity to that obtained when using the average velocities of systems nearby \sys{}'s projection onto the Galactic mid-plane using GAIA \citep{GAIADR3}, and find no significant quantitative difference. We also reran the analysis using \texttt{gala}'s \texttt{MilkyWayPotential} \citep{gala,apwzenodo,Bovy2015} and found that our results are not significantly affected by our particular choice of Galactic potential.

\subsection{Binary Evolution}
\label{sec:evolution}

\begin{figure}
\includegraphics[width=0.45\textwidth]{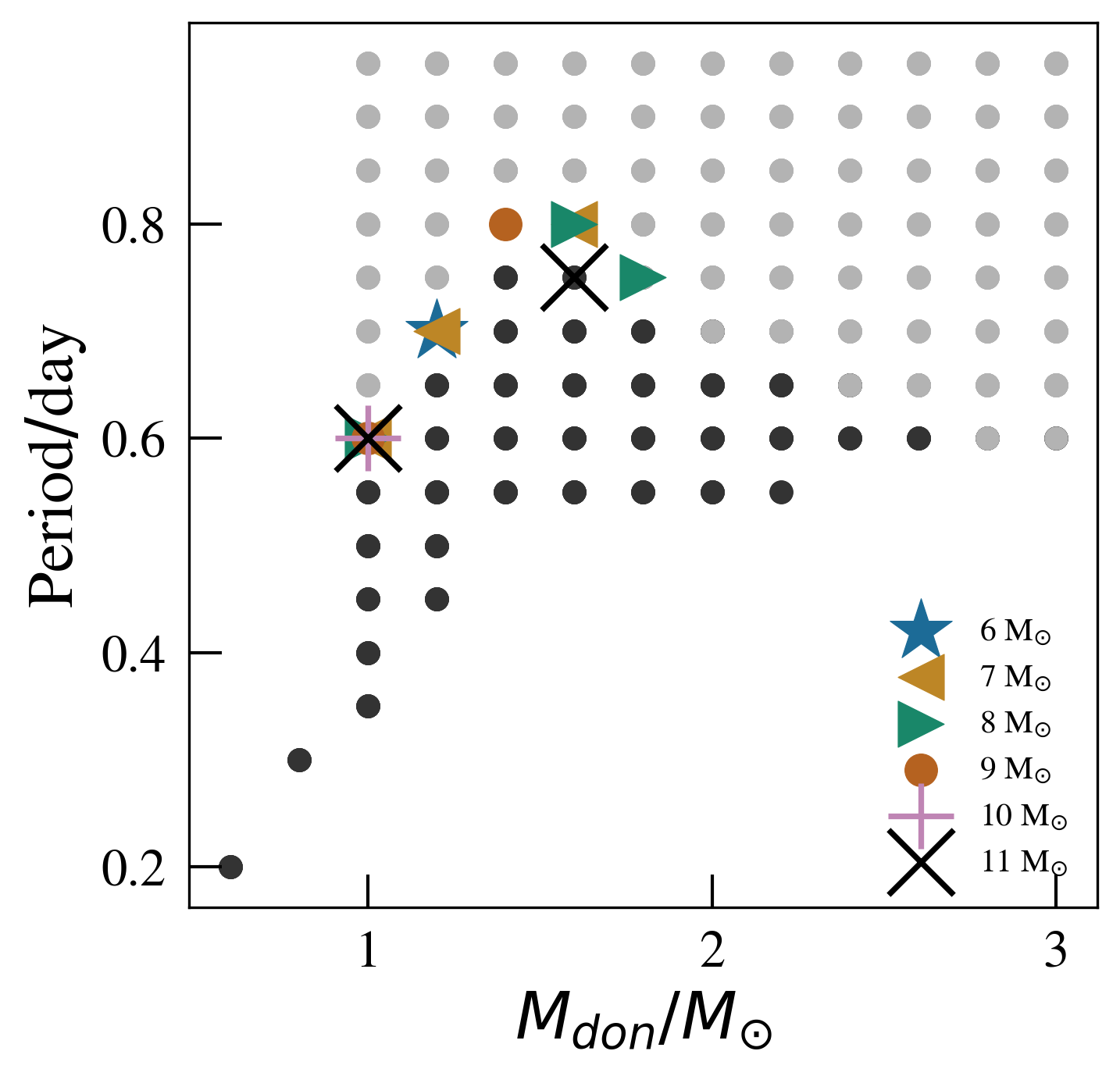}
\caption{The viable RLO2 parameter space explored with \texttt{MESA}. The colored shapes mark successful binaries (at different BH masses) that evolved to simultaneously satisfy all observational constraints on the BH and donor masses, donor effective temperature, and orbital period of \sys{} to within 2-$\sigma$. In black are unsuccessful binaries where magnetic braking was efficient, shrinking their periods throughout mass transfer. In grey are unsuccessful binaries above the bifurcation period, expanding during mass transfer.  }
\label{fig:winners}
\end{figure}
\begin{figure}
\includegraphics[width=0.45\textwidth]{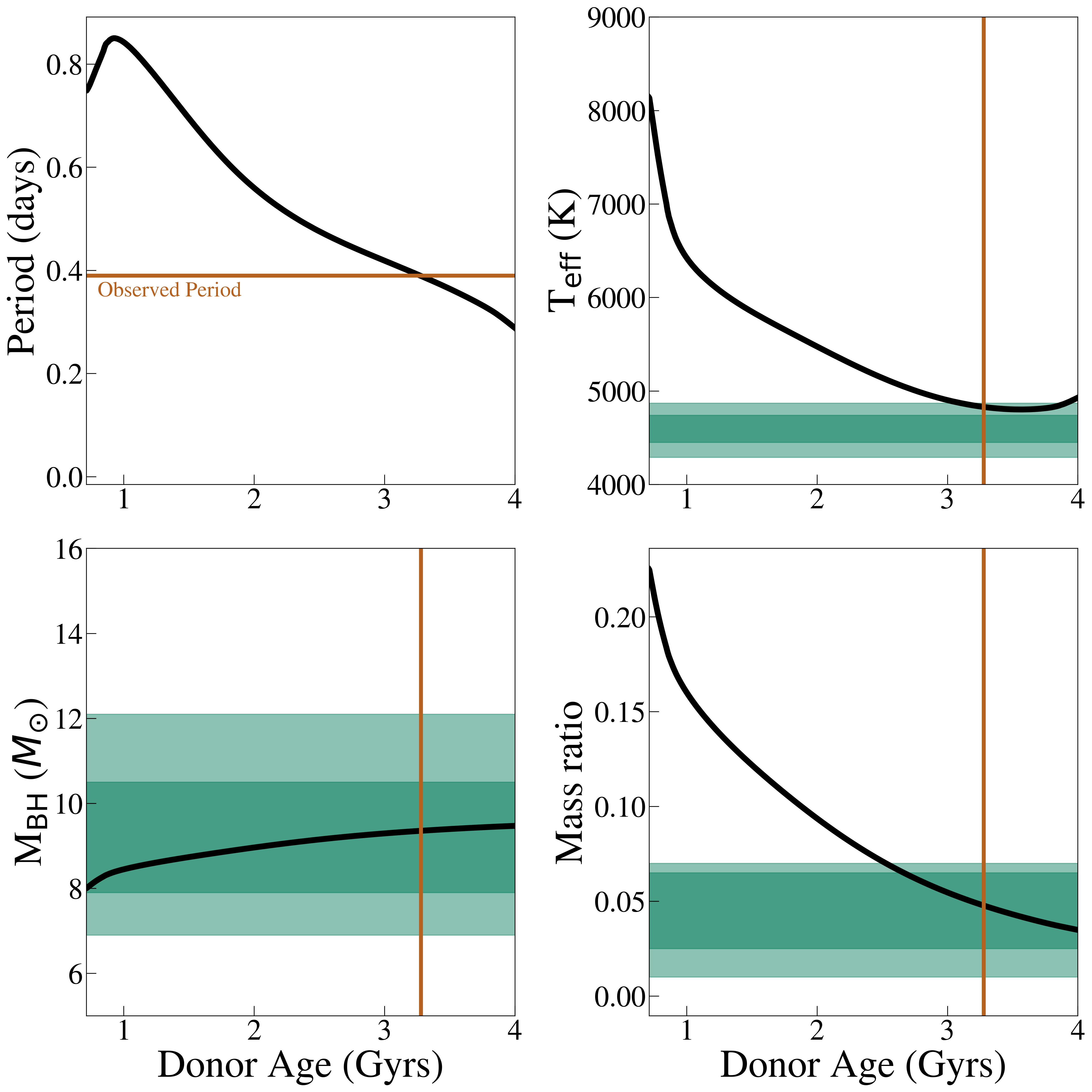}
\caption{Example of a successful RLO2 system. The 1.8 \Msun donor fills its Roche lobe at a period of 0.75 days with an 8 \Msun{} BH. The orange line denotes the age of the donor at the observed period. The green shaded regions mark the 2-$\sigma$ confidence intervals on the observed black hole mass, mass ratio, and donor effective temperature.}
\label{fig:sequence}
\end{figure}
\begin{figure} 
\includegraphics[width=0.45\textwidth]{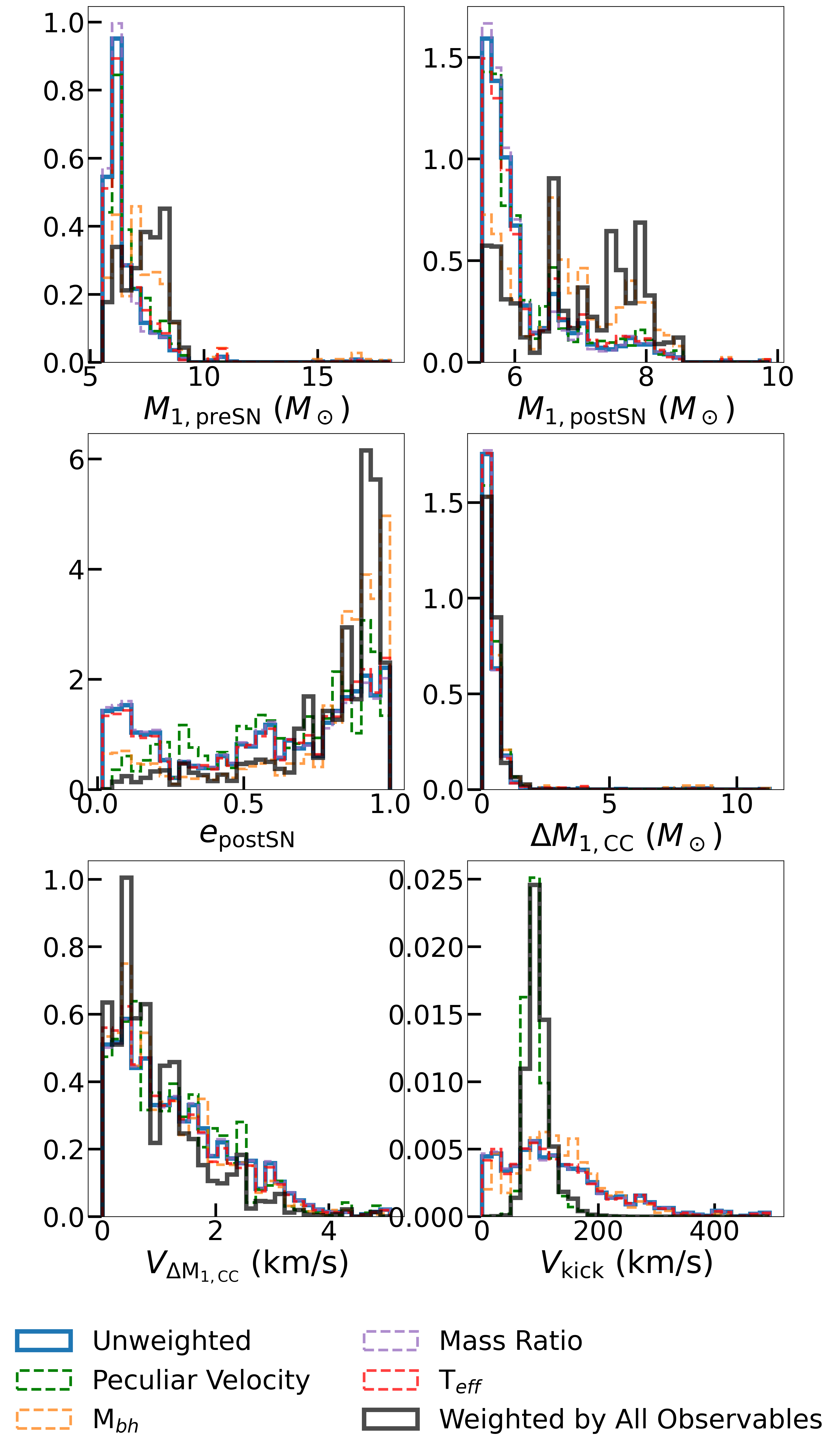}
\caption{Posteriors over parameters and kicks of the potential progenitor binaries \chase{just before and after SN.} In blue we plot the unweighted posteriors. In green, orange, purple, and red we weight these posteriors individually according to the observed \VpecBirthObs, \MbhObs, \qObs, \TeffObs. In black we plot our final result, weighted by the product of the individual weights. }
\label{fig:progenitor}
\end{figure}
\chase{We now use the possible pre- and post-SN parameter space of \sys{}'s progenitor in order to extract the BH natal kick from the total imparted systemic velocity inferred from its kinematics. We do this in two steps:}
First we constrain the BH mass, donor mass, and period (\MbhRLO, \MdonRLO, \PRLO) at the onset of mass transfer from the secondary onto the black hole. We consider a successful RLO2 system to be any binary that evolves to simultaneously satisfy observational constraints on the BH and donor masses, donor effective temperature, and orbital period of \sys{} to within 2-$\sigma$ of their observed values. We then use \posydon{} to evolve a population of ZAMS binaries up until RLO2 while modeling  and sampling over natal kick velocities, matching the results to our successful RLO2 systems and identifying potential \sys{} progenitors. 

The mass of the black hole at RLO2 is constrained from above by the upper bound on the observed black hole mass M$_{bh,obs}\lesssim \MbhHigh$ \Msun. Meanwhile, the RLO2 period is constrained from below by the requirement that the donor is not already filling its Roche lobe at ZAMS, and from above by the requirement that it will fill its Roche lobe in a Hubble time. With the exception of very low donor masses \MdonRLO$\lesssim0.75$ \Msun, this lower bound means that \sys's orbit must have \textit{shrunk} since RLO2 \footnote{While it is possible that \sys{} entered RLO2 immediately after a CE, or immediately post-SN due to a fortuitous kick, this becomes less likely when accounting for observations that find that the donor is evolved. Due to the high mass of the BH progenitor relative to the donor, this would make it unlikely that the donor began mass transfer at such an early stage given we are still observing it with such a short period today.}. Since subsequent mass transfer from the donor onto the BH would \textit{expand} \sys's orbit ($q<1$), this is only possible with efficient magnetic braking. \chase{Assuming the companion stays tidally locked, spin angular momentum lost through magnetic braking would be compensated by the orbital angular momentum of the binary, shrinking its orbit.} Since magnetic braking is inefficient in massive stars and at large periods, this adds the constraint that \MdonRLO{} is sufficiently small so that by the time it loses enough mass to enter the regime where magnetic braking is efficient, the period is still below the bifurcation period that delineates whether mass transfer will grow or shrink the orbit in the presence of magnetic braking. We find that this effectively limits $\MdonRLO<3$ \Msun{} and $\PRLO<1$ day.

From these constraints, we construct a grid of binaries -- spaced uniformly in BH mass, donor mass, and orbital period at intervals of 1 \Msun, 0.2 \Msun, and 0.05 days, respectively -- for which we calculate detailed mass transfer sequences using \texttt{MESA}. Figure \ref{fig:winners} shows results for selected slices of that grid. Of these sequences, we find 12 -- marked in colored shapes in Figure \ref{fig:winners} -- that simultaneously satisfy observational constraints on the masses, period, and effective donor temperature of \sys{} within 2-$\sigma$. We refer to these as successful RLO2 binaries. We plot unsuccessful binaries that expanded throughout mass transfer with grey circles, and those below the bifurcation period that shrank due to efficient magnetic braking with black circles. In Figure \ref{fig:sequence}, we plot an example of a successful RLO2 mass transfer sequence.

Having identified the space of potential \sys{} progenitors at RLO2, we can continue to narrow down the potential space of \sys{} progenitors at ZAMS. We begin by evolving a nominal population of 10,000,000 ZAMS binaries via \posydon{}'s interpolation scheme over the \texttt{MESA} grids described in \citet{Posydon}. We choose a population flat in mass ratio, with orbital separation sampled log-uniform between 5 and 10$^4$ R$_{\odot}$, and primary mass \MbhZams{} drawn from a Kroupa IMF \citep{Kroupa2001} between 7 and 120 \Msun. It is reasonable to assume that \sys{} did not undergo mass inversion and that $M_2$ will lose a negligible amount of mass between ZAMS and RLO2 -- consistent with predictions of negligible winds in low-mass stars -- so we limit $\MdonZams<3$ \Msun{} by the above argument that requires progenitors at RLO2 to be in a regime where magnetic braking is efficient. We evolve each binary until the core-collapse of the primary (CC1), discarding systems that merged before compact object formation. Again assuming that $M_2$ will lose negligible mass before RLO2, we also discard any systems where the pre-SN donor mass $\MdonPre>3$ \Msun, leaving a population of $\sim$30,000 pre-SN binaries that could viably be progenitors to \sys{}.

Finally we evolve the viable pre-SN binaries through the core collapse of the primary and the subsequent detached binary evolution up until RLO2. Rather than assume any particular supernova prescription, we take a broad, agnostic approach to modeling the core collapse, sampling mass loss uniformly such that the remnant black hole mass is between the minimum and maximum of our winning RLO black hole masses, and natal kicks \Vkick{} uniformly between 0 and 500 km/s. We repeat this sampling 6,000 times per binary, calculating the post-supernova orbital properties and systemic velocities according to \citet{Kalogera1996}, and evolve the resulting BH + H-rich star binaries until the donor fills its Roche lobe at RLO2. We then identify potential \sys{} progenitors from this population by matching them to the cells of our successful RLO2 binaries, forming $\sim$850 potential evolutionary histories for \sys{}. We refer to these binaries as ``potential progenitors". \chase{We note that by sampling uniformly over mass loss and kick velocity at SN, our span of potential progenitors is conservative.} We discuss our results and how we weigh them by the observed properties of \sys{} in Section \ref{sec:results}.

\subsection{Constraints on the natal BH kick and progenitor properties}
\label{sec:results}

\begin{figure}
\includegraphics[width=0.45\textwidth]{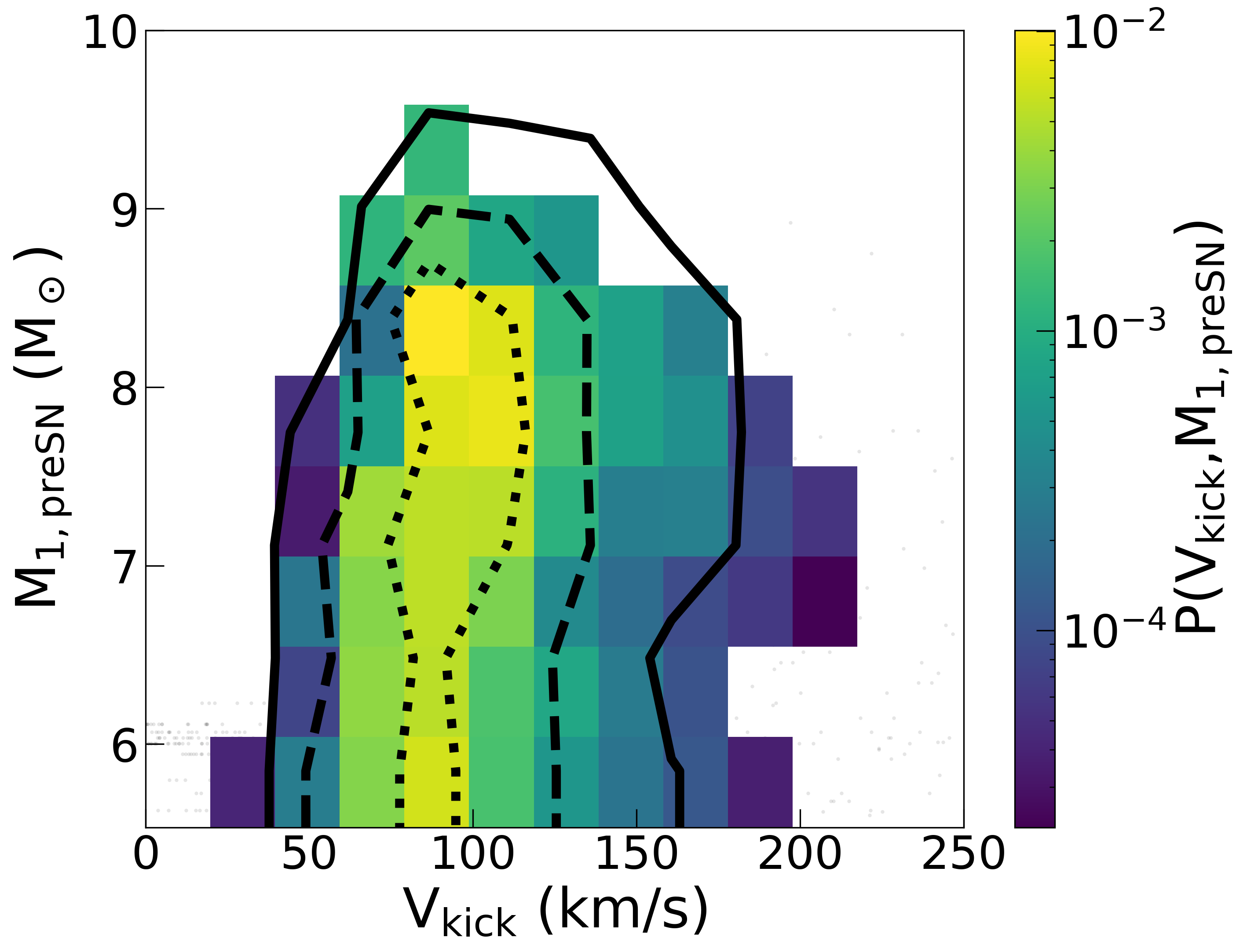}
\caption{Here we plot the final marginalized posterior over pre-SN primary mass and kick velocity, with the final, combined weights. The dotted, dashed, and solid contours enclose central 68\%, 90\%, and 99\% probability.}
\label{fig:MpreVkick}
\end{figure}
Since not all potential progenitors to \sys{} are equally successful, we weigh the distributions over our potential progenitor parameters by the observed properties of \sys{}. That is, given a set of observable quantities $\vec{\theta}_f$ at present-day corresponding to a progenitor binary with parameters $\vec{\theta}_i$, $p(\vec{\theta}_i)$ is weighted by the probability of the observable quantities given the electromagnetic data EM:
\begin{equation}
    p(\vec{\theta}_i |\mathrm{EM}) \propto p(\vec{\theta}_i)p(\vec{\theta}_f|\mathrm{EM}).
\end{equation}
For $\vec{\theta}_f \in \{\MbhObs, \qObs, \TeffObs\}$ we use (asymmetric as appropriate) Gaussians to fit the corresponding posteriors reported in \citet{Sanchez2021}. For \VpecBirthObs, we use a kernel density estimation to fit the posterior shown in the bottom-right panel of Figure \ref{fig:kinematics}. We plot the results in Figure \ref{fig:progenitor}. In blue we plot the unweighted posteriors, which is the set of binaries that at some point had a BH mass, mass ratio, and effective temperature within 2$\sigma$ of the observed value. In green, orange, purple, and red we weight these posteriors individually according to the observed \VpecBirthObs, \MbhObs, \qObs, \TeffObs. Our final results, in black, are obtained by multiplying all of the observational weights together. In Figure 5, we plot the final marginalized posterior over the pre-SN primary mass and kick velocity, weighted by all observations.

    
    Although we allowed for arbitrarily high mass loss upon core collapse, 
we find that $\Delta\mathrm{M}_\mathrm{1,CC} < \dMHighwAll$ \Msun{} with 95\% confidence. With low $\Delta \mathrm{M}_\mathrm{1,CC}$ the velocity imparted on the center of mass via mass loss alone would have been small, with $\mathrm{V}_{\Delta \mathrm{M}} = \VblaauwMedwAll_{-\VblaauwMinuswAll}^{+\VblaauwPluswAll}$ km/s. In the absence of an additional SN kick on the remnant BH, the systemic velocities would be incompatible with the inferred \VpecBirthObs. Indeed, once weighted by observations \chase{-- and crucially the kinematic constraints on the total birth velocity --} we find that $\Vkick > \VkickLowwAll$ km/s with 95\% confidence, with $\Vkick = \VkickMedwAll_{-\VkickMinuswAll}^{+\VkickPluswAll}$ km/s. 

The presence of an SN kick may also have helped \sys{} reach the short periods at RLO2 that we find to be necessary to produce it. Since the progenitor would have circularized during a common envelope phase, the post-SN orbit would have mild eccentricity -- and therefore would have difficulty shedding angular momentum -- in the absence of a kick and with little to no mass loss. Indeed, we find that all but the shortest-period pre-SN potential progenitors received a kick. Before weighting by the inferred \VpecBirth, we find that of the progenitors that received a kick of $<10$ km/s, all had pre-SN periods of $<$5 days, and none had post-SN eccentricities of $>$ 0.1.

\chase{We stress that these results should be understood as constraints inferred using flat priors, and are therefore conservative. Our finding of a scenario with low mass-loss and non-negligible natal kick \result{are at odds} with SN prescriptions that scale BH kicks by mass-loss \citep{Bray2018, Giacobbo2020} \result{resulting in negligible kicks with low mass-loss}.  On the other hand, simulations in \citet{Coleman2022}\result{ -- listed in Table 1 and discussed in Section 3.1.3 therein -- find that BHs may receive natal kicks of up to $\simeq$ 75} km/s due to relativistic asymmetric neutrino emission alone, with negligible contribution from matter ejecta.}

\section{Conclusions}
\label{sec:conclusions}
We find that \system{} is only the second BH system that -- when analyzed in the context of \textit{both} its kinematic and binary/stellar history \chase{under the assumption that it was formed in the Galactic disk via isolated binary evolution, and assuming uninformative priors on the mass-loss and natal kick velocity imparted at SN}-- requires a natal kick to exist. Our inferred natal kick of $\VkickMedwAll_{-\VkickMinuswAll}^{+\VkickPluswAll}$ km/s is consistent with \citet{Sanchez2021}, who estimated from kinematics alone that the total center-of-mass velocity imparted to this system was $75_{-12}^{+25}$ km/s. It is also consistent with \citet{Atri}, who found that the population of BH LMXB birth velocities is well fit by a normal distribution peaking around 107 km/s with a standard deviation of 16 km/s. Our result suggests that the birth velocity of \sys{} could not have been provided by symmetric mass-loss alone, but requires an additional natal kick. Our 95\% lower limit on this natal kick of \VkickLowwAll{} km/s is also similar to the findings of  \citet{Fragos2009} for XTE J1118+40, who found that its BH received a kick of at least $\simeq$\,80 km/s. With this analysis of \sys{}, we strengthen the case that at least some BHs receive natal kicks at birth. 

We find that our results are robust against kinematic uncertainties. The inferred boost to the center of mass velocity from symmetric mass loss alone, $\mathrm{V}_{\Delta \mathrm{M}} = \VblaauwMedwAll_{-\VblaauwMinuswAll}^{+\VblaauwPluswAll}$ km/s, would be incompatible with even moderate birth peculiar velocities $\gtrsim$ 10 km/s. We assess that our inference is not qualitatively affected by our choice of Milky Way potential nor the assumption that \sys{} was born exactly in the Galactic mid-plane with a circular velocity. Further, as shown in \citet{Fragos2009} for a system with a very similar donor, the kick constraints are not sensitive to the assumed magnetic braking law. We do note that, as in all binary evolution calculations to date we assume that binaries instantly circularize upon the onset of mass transfer. This is not always a well-justified assumption and it leads to a mismatch between the donor ages produced in our population synthesis and in our individual \texttt{MESA} runs that matched the observed properties; it is possible that correcting for this effect (Rocha et al. 2022, in prep.) may have a small quantitative effect on our kick constraints. Lastly, we note that at $\sim 1$ kpc above the Galactic mid-plane, we cannot exclude that \sys{} formed dynamically before being ejected from a globular cluster, in which case our analysis is not applicable. 

Through this and other past studies we have demonstrated that accounting for all observational characteristics and coupling binary evolution and kinematic modeling we can provide robust constraints on natal BH kicks and associated mass loss at formation. With an increasing sample of LMXBs with proper motion and radial velocity estimates, our goal in future work is to investigate potential statistical correlations between BH kicks and mass loss characteristics at birth, which may shed light on the natal kick mechanism. 

\section*{Acknowledgements}
We thank Christopher Berry, Lieke van Son, and Michael Zevin for helpful discussions and comments on the manuscript. CK is supported by the Riedel Family Fellowship. VK is grateful for support from a Guggenheim Fellowship, from CIFAR as a Senior Fellow, and from Northwestern University, including the Daniel I. Linzer Distinguished University Professorship fund. KR and DM thank the LSSTC Data Science Fellowship Program, which is funded by the LSST Corporation, NSF Cybertraining Grant No. 1829740, the Brinson Foundation, and the Gordon and Betty Moore Foundation; their participation in the program has benefited this work. ZD is grateful for support from the CIERA Board of Visitors Research Professorship. KK and EZ were partially supported by the Federal Commission for Scholarships for Foreign Students for the Swiss Government Excellence Scholarship (ESKAS No.\ 2021.0277 and ESKAS No.\ 2019.0091, respectively). ZX was supported by the Chinese Scholarship Council (CSC). This work was performed with the help of the computing resources at CIERA provided by the Quest high performance computing facility at Northwestern University -- funded through NSF PHY-1726951 -- which is jointly supported by the Office of the Provost, the Office for Research, and Northwestern University Information Technology.

\software{\texttt{NumPy} \citep{numpy},\,
\texttt{SciPy} \citep{scipy},\,
 \texttt{matplotlib} \citep{matplotlib},\,
 \texttt{pandas} \citep{pandas},\,
 \texttt{POSYDON} \citep{Posydon}}

\bibliography{references.bib}
\end{document}